\documentclass[sigconf]{acmart}

\AtBeginDocument{%
  }

\setcopyright{acmlicensed}
\copyrightyear{2018}
\acmYear{2018}
\acmDOI{XXXXXXX.XXXXXXX}

\acmConference[Conference acronym 'XX]{Make sure to enter the correct
  conference title from your rights confirmation emai}{June 03--05,
  2018}{Woodstock, NY}
\acmISBN{978-1-4503-XXXX-X/18/06}

\usepackage{pifont}

\newcommand{\cmark}{\textcolor{green!60!black}{\ding{51}}}  % checkmark
\newcommand{\xmark}{\textcolor{red}{\ding{55}}}    
\begin{document}

\title{WiReSens Toolkit: An Open-source Platform towards Accessible Wireless Tactile Sensing}

\author{Devin Murphy}
\affiliation{%
  \institution{MIT CSAIL}
  \city{Cambridge}
  \country{USA}
}
\email{devinmur@mit.edu}

\author{Junyi Zhu}
\affiliation{%
  \institution{University of Michigan EECS}
  \city{Ann Arbor}
  \country{USA}
}
\email{zhujunyi@umich.edu}

\author{Paul Pu Liang}
\affiliation{%
  \institution{MIT Media Lab and EECS}
  \city{Cambridge}
  \country{USA}
}
\email{ppliang@mit.edu}

\author{Yiyue Luo}
\affiliation{%
  \institution{University of Washington ECE}
  \city{Seattle}
  \country{USA}
}
\email{yiyueluo@uw.edu}

\author{Wojciech Matusik}
\affiliation{%
  \institution{MIT CSAIL}
  \city{Cambridge}
  \country{USA}
}
\email{wojciech@csail.mit.edu}

\renewcommand{\shortauthors}{Murphy et. al}

\begin{abstract}
Past research has widely explored the design and fabrication of resistive matrix-based tactile sensors as a means of creating touch-sensitive devices. However, developing portable, adaptive, and long-lasting tactile sensing systems that incorporate these sensors remains challenging for individuals having limited prior experience with them. To address this, we developed the WiReSens Toolkit, an open-source platform for accessible wireless tactile sensing. Central to our approach is adaptive hardware for interfacing with resistive sensors and a web-based GUI that mediates access to complex functionalities for developing scalable tactile sensing systems, including 1) multi-device programming and wireless visualization across three distinct communication protocols 2) autocalibration methods for adaptive sensitivity and 3) intermittent data transmission for low-power operation. We validated the toolkit's usability through a user study with 11 novice participants, who, on average, successfully configured a tactile sensor with over 95\% accuracy in under five minutes, calibrated sensors 10× faster than baseline methods, and demonstrated enhanced tactile data sense-making.
\end{abstract}

%%Technical evaluations demonstrate scalability up to five simultaneous devices, adaptivity for a variety of sensors and pressure ranges, and up to a 42\% increase in device lifetime.  A user study further highlights the toolkit's excellent usability through its support for rapid system configuration, automatic calibration, and physically-situated visualization. We demonstrate our platform's flexibility by using it to prototype applications such as musical gloves, gait monitoring shoe soles, and IoT-enabled smart home systems, and end with a discussion on 





\begin{teaserfigure}
  \includegraphics[width=\linewidth]{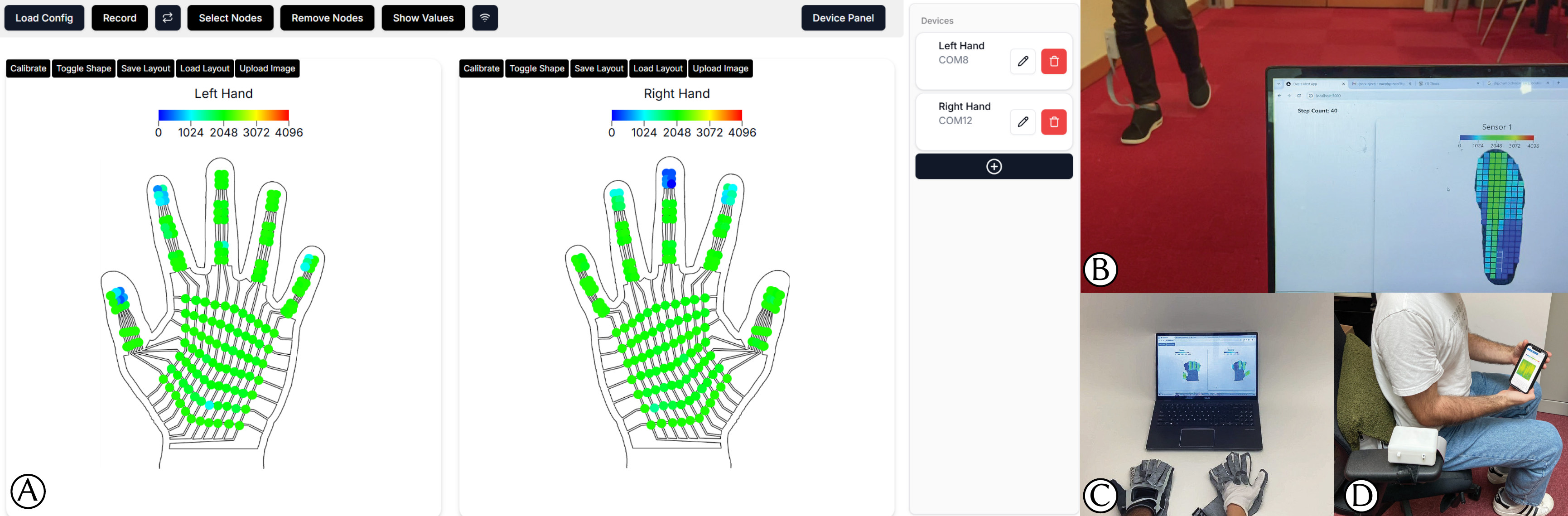}
    \caption{WiReSens Toolkit provides open-source hardware and software to enable the development of portable, adaptive, and efficient resistive tactile sensing systems.}
  \label{fig:teaser}
\end{teaserfigure}

\maketitle

\section{Introduction}
Tactile sensors endow humans and everyday objects with enhanced ability to capture, understand, and augment physical  interactions, increasingly finding applications within cyber-physical systems \cite{2017MalvadePressure}, the Internet of Things (IoT) \cite{Anwer2022}, and Human-Computer Interaction (HCI) \citep{TactileKeyboard, affectiveTouch2024}. While a variety of principles have been employed to sense touch, many researchers adopt resistive matrix-based approaches \cite{sundaram2019learning, luo2021learning, wicaksono20223dknits} where a piezoresistive material is sandwiched between two orthogonal electrode arrays to measure applied pressure via resistance changes. This approach is appealing because it is intuitive,  reduces the need for extensive wiring \cite{luo2021intelligent}, and past research has demonstrated that it can be implemented via a variety of fabrication techniques \citep{wicaksono20223dknits,AignerEmbroidery2020,3dprinted2024,murphy2025fits}.

To realize a vision of ubiquitous tactile sensing for large-scale understanding of human-environment interactions, infrastructure beyond sensor manufacturing is essential --- researchers need robust tools for system deployment. Large scale sensing requires consideration of additional factors that presently make this deployment difficult. Consider a pair of tactile sensing gloves designed to monitor bimanual manipulation tasks. Because the gloves are to be worn, they must transmit data wirelessly, which immediately raises the question: which protocol to use? Bluetooth? Wi-Fi? And how should data be synchronized across both hands? This challenge compounds when battery life is taken into account—wireless transmission drains power quickly \cite{Soua2011WSN}, limiting the duration over which interactions can be captured unless the user implements custom energy-saving strategies. Then there’s the issue of calibration: resistive sensors output different ranges of signals depending on how they are fabricated and may be used to capture everything from a light graze to a forceful impact. So how does one ensure consistent pressure readouts when the gloves are used by different people, in different application contexts, or re-fabricated for a new study? Without robust infrastructure for handling these issues, scaling up to broader deployments or adapting to new applications becomes tedious at best and impractical at worst. 

To address these challenges, this paper presents the WiReSens Toolkit (Fig. \ref{fig:teaser}). Inspired by sensing toolkits that simplify prototyping through domain-specific abstractions \citep{zhu2021eit,savage2012midas,firmwareWearableKit}, our platform consists of a web-based GUI that interfaces with open-source hardware, making it easier to develop portable, adaptive, and efficient resistive tactile sensing systems. It includes abstractions for sensor array readout, wireless communication via three distinct protocols (Wi-Fi, Bluetooth Low Energy (BLE), and ESP-NOW), real-time interactive data visualizations, automatic sensitivity calibration,  and intermittent data transmission for energy savings, all of which are lacking in existing open-source solutions for tactile sensing \citep{mdonneaud2017Matrix, pourjafarian2019multi, arduino_capacitivesensor}.

Our technical evaluation of WiReSens Toolkit evidences its ability to provide robust multi-sender wireless communication across all three protocols, optimized pressure resolution during sensor readout, and up to a 42\% increase in tactile sensing device lifetime. We also conducted a user study with 11 novice users, which demonstrated that the platform has excellent usability, streamlines access to complex functionality, and helps users better interpret pressure data visualizations. We finally demonstrate the flexibility of our toolkit by using it to rapidly prototype various applications, including musical gloves, shoe soles for gait monitoring, a pillow for posture monitoring and remote control, and a smart home welcome mat. 

In summary, this work contributes:

\begin{itemize}
    \item A web GUI for \textbf{quick reconfiguration} of resistive tactile sensing devices and real-time \textbf{interactive data visualization}.
    \item An open-source resistive sensor readout circuit in two sizes, with an additional \textbf{adaptive module} and on-device \textbf{auto-calibration methods} that allow users to tune the sensitivity of devices.
    \item Methods for \textbf{interaction-aware wireless data transmission} via intermittent sending, to \textbf{save power} during periods of tactile sensor inactivity.
    \item \textbf{A technical evaluation} demonstrating robust wireless communication across multiple protocols, adaptive readout optimization for improved pressure resolution, and up to a 42\% increase in device lifetime. 
    \item \textbf{A user study} highlighting the toolkit's usability during device configuration, calibration, and visualization tasks.
    \item \textbf{Example applications} to show how the toolkit facilitates rapid system deployment in various contexts. 
\end{itemize}

\section{Related Work}

\begin{table*}[h!]
\centering
% \caption{Comparison of Capacitive and Resistive Sensing Toolkits}
\resizebox{\linewidth}{!}{%
\begin{tabular}{@{}lcccccc@{}}
\toprule
 & \textbf{CapToolKit \cite{wimmer2007capacitive}} & \textbf{Midas \cite{savage2012midas}} & \textbf{Multi-Touch Kit \cite{pourjafarian2019multi}} & \textbf{E256 \cite{mdonneaud2017Matrix}} & \textbf{3D ViTac} \cite{huang20243dvitac} & \textbf{Ours} \\ 
 \midrule
\textbf{Sensing principle} & Capacitive & Capacitive & Capacitive & Resistive & Resistive & Resistive \\
\textbf{\# of sensors per device} & 8 & 9 & 256 & 256 & 256 & \textbf{1024} \\
\textbf{\# of sending devices} & 1 & 1 & 1 & 1 & 4 & \textbf{5} \\
\textbf{Sensitivity calibration} & \xmark & \xmark & \xmark & \xmark & \xmark & \cmark \\
\textbf{Wireless protocol support} & \xmark & \xmark & \xmark & \xmark & \xmark & \cmark \\
\textbf{Power-saving mode} & \xmark & \xmark & \xmark & \xmark & \xmark & \cmark \\
\textbf{Visualization} & Static & \textbf{Configurable} & Static & Static & Static & \textbf{Configurable} \\
\bottomrule
\end{tabular}%
}
\vspace{0.2em}
\caption{In contrast to existing open-source tactile sensing toolkits, WiReSens enables sensing on a larger scale (4x the number of sensors and 5x the number of devices), automatic sensitivity calibration, wireless protocol support, and power-saving operation modes.}
\label{table:priorworks}
\end{table*}

The WiReSens Toolkit is built upon the principle of resistive pressure sensing and is most similar to prior works which facilitate the creation of interactive touch-sensing systems. We first provide an overview of resistive pressure sensing before situating WiReSens Toolkit within the landscape of toolkits designed for tactile sensing.

\subsection{Resistive Pressure Sensing}
Past research has explored various methods for dynamic pressure sensing, with optical, capacitive, and resistive techniques being among the most widely used. Optical tactile sensors, such as GelSight \cite{2017YuanGelsight}, analyze light pattern changes caused by material deformation under pressure. While offering high resolution and performance, these sensors often rely on proprietary hardware and face limitations in scalability and flexibility due to visual occlusion challenges. Capacitive sensing, on the other hand, typically involves two conductive electrodes separated by a deformable dielectric layer, where applied pressure alters capacitance \cite{Li2021CapPrinciple}. However, capacitive methods often require complex readout circuits \citep{Vu2021, CHEN2020175} or shielding layers to mitigate electromagnetic noise \cite{Srinivas2024CapEMI}. Additionally, capacitive sensing is more commonly used in a mutual \citep{commonground2017}  or swept-frequency \cite{touche2012} setup for binary touch detection rather than precise pressure measurement. This work therefore focuses on resistive-based pressure sensing.

\begin{figure}
    \includegraphics[width=0.8\linewidth]{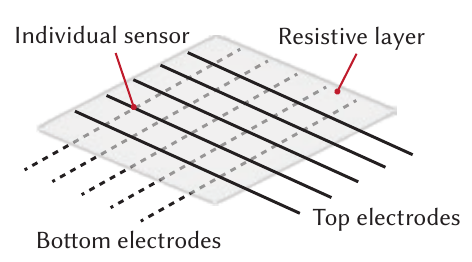}
  \vspace{-1em}
  \caption{Typical layout of a resistive matrix-based pressure sensing form factor, with a resistive layer sandwiched between two sets of orthogonally positioned electrodes.}
  \label{fig:design}
\end{figure}

In resistive pressure sensing, orthogonally arranged electrodes sandwiching a piezoresistive layer form a 2D sensing matrix, where pressure is measured via resistance changes at electrode intersections \cite{BIJENDER2022CapResPressure} (Fig. \ref{fig:design}). This approach avoids the scalability problem that is present in optical pressure sensors, as wires can easily be added or removed and spaced tighter or more spread out. When combined with digital fabrication approaches like machine knitting \cite{wicaksono20223dknits}, digital embroidery \cite{AignerEmbroidery2020}, 3d printing \cite{3dprinted2024}, and flexible circuit printing \cite{murphy2025fits}, it has allowed resistive sensing to be incorporated in a variety of form factors like hand bags \cite{2018REsiParzer}, carpets \citep{luo2021learning, Wicaksono2022Tapis}, and even hair extensions \cite{FeatherHair2022}. We show how our toolkit is adaptable to these various fabrication approaches in this paper's \textit{Toolkit Evaluation} section. 

Pressure signals are typically read from a resistive sensing matrix using a zero potential scanning readout circuit \citep{Kim2016Scanning, DALESSIO199971}, the hardware on which WiReSens Toolkit is also built (Fig. \ref{fig:circuit}B). The circuit topology is well established and widely adopted due to its ability to reduce crosstalk between neighboring electrodes \cite{DALESSIO199971}. Although some works provide similar open-source hardware \citep{huang20243dvitac,  mdonneaud2017Matrix, fiedler2022low}, WiReSens Toolkit supports a higher density (up to 1024 pressure sensing locations), and to our knowledge is the first open-source hardware which includes a digital potentiometer in this architecture for tuning sensitivity.

\subsection{Touch Sensing Toolkits}
A range of toolkits have been developed to facilitate prototyping touch sensing applications, with early works focusing primarily on capacitive sensing. CapToolKit by \citet{wimmer2007capacitive} and Midas by \citet{savage2012midas} were among the first to provide open-source hardware and software to streamline the prototyping of capacitive touch interfaces. Midas lowered the barrier to designing custom capacitive sensors which could be fabricated using digital manufacturing techniques such as inkjet printing, vinyl cutting, and CNC milling. These contributions allowed researchers and designers to rapidly iterate on new touch-sensing applications without needing deep expertise in circuit design or signal processing. Building on this foundation, the Multi-Touch Kit \cite{pourjafarian2019multi} further abstracted capacitive sensor processing by introducing an Arduino library for mutual capacitance-based multi-touch arrays. This allowed capacitive touch sensors to be implemented using only a standard microcontroller (MCU). While these toolkits provided valuable abstractions for capacitive touch sensing, they primarily supported binary multi-touch detection.

Support for resistive touch sensing is typically limited to specific application contexts and sensor types. The E256 project by \citet{mdonneaud2017Matrix} provides open-source Arduino firmware for denoised readout of resistive matrix-based eTextiles designed for musical expression. By leveraging a fabric-based piezoresistive sensor, E256 enables pressure-sensitive interaction beyond simple binary touch detection. Similar to the Multi-Touch kit, it includes Processing software for real-time visualization of sensor readings. Additionally, the 3D ViTac system by \citet{huang20243dvitac} has expanded accessibility within robotics communities by offering a hardware guide, Arduino firmware, and Python ROS libraries for parallel gripper resistive tactile sensors.

WiReSens Toolkit builds upon these prior works (Table \ref{table:priorworks}) across key areas of HCI toolkit contributions as identified by \citet{Ledo2018Toolkits} and our design goals of portability, adaptivity, and efficiency. With built-in wireless communication support for up to five devices, each capable of sensing dynamic pressure at 1024 unique locations, WiReSens Toolkit extends beyond individual sensors to enable the development of distributed multi-sensor interactive systems. This allows novice users to explore tactile interaction design on a larger scale. By integrating features such as automatic sensitvity calibration and power-saving modes, it simplifies access to more complex functionality, enabling users to rapidly prototype tactile sensing systems which are adaptable and sustainable in addition to functional. Finally, WiReSens is the first of these toolkits to be evaluated with first-time users, demonstrating that even those without prior experience can set up a functional tactile sensing system in 5 minutes.

\section{Design Requirements}
To ensure the WiReSens Toolkit would be effective for developing scalable tactile sensing systems, we outlined several key goals for its design.

\paragraph{\textbf{Portability:}} Tactile sensing devices must be easily repositioned to capture interactions across different surfaces and environments. This means wireless communication is a necessity, especially in wearable applications where wires can restrict movement. While many sensing toolkits adopt a single communication protocol (often Bluetooth \citep{zhu2021eit,Dementyev2015SensorTape, voxelhap2023}), different protocols offer distinct advantages \cite{commStudy}. For example, Wi-Fi offers excellent speed and reliability indoors but is difficult to access outdoors, where Bluetooth communication is more viable. A toolkit that simplifies switching between protocols would broaden the range of environments where tactile sensing systems can be effectively implemented.

\paragraph{\textbf{Adaptivity:}} A key challenge with resistive tactile sensors is that the resistance-force relationship varies across fabrication techniques, and different applications require distinct force sensitivity ranges. In these different contexts, the same readout circuit can either overamplify or underamplify signals, limiting pressure resolution \cite{RSAComparison}. Additionally, resistive sensing array data is typically displayed statically, with each sensing node fixed in place. However, the real-world positioning of these nodes can shift when different users wear the same sensor or as conditions change over time. An adaptive visualization that accounts for these shifts could help users interpret tactile interactions more accurately and intuitively.

\paragraph{\textbf{Efficiency:}} Wireless communication is power-intensive, which limits device lifetime and poses challenges for sustainable development \cite{Soua2011WSN,ZhaoIoTEfficiency2019}. Low-power operation is therefore important to enable large-scale data collection over extended periods of time, though it can affect the computational capabilities of devices. Users should have access to methods that allow them to navigate this trade-off. Additionally, users may invest significant time and resources into developing interactive systems, so reconfiguring them for different form factors, users, and applications should be seamless, minimizing hardware and firmware modifications.

\begin{figure*}
  \includegraphics[width=\linewidth]{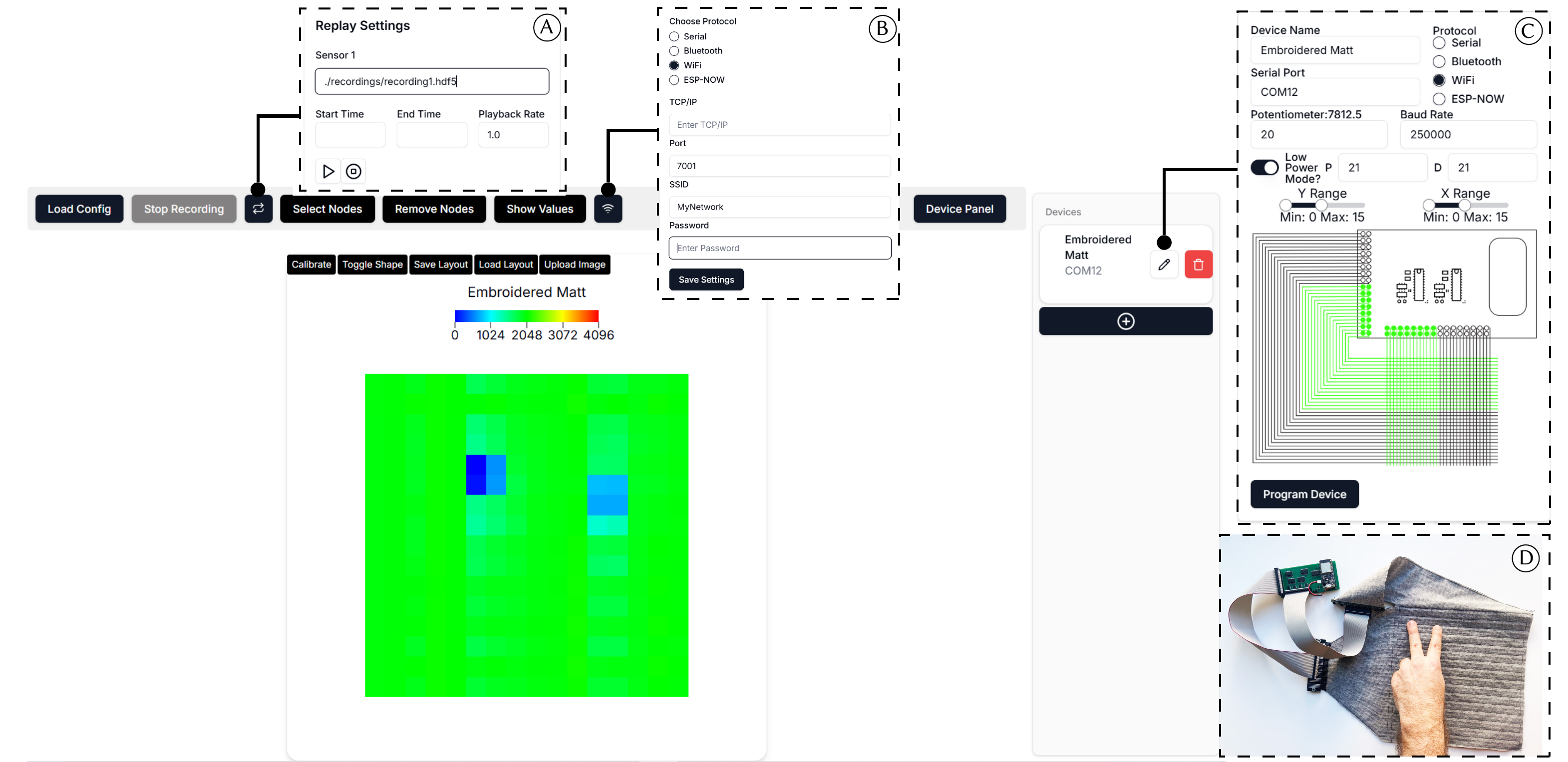}
  \vspace{-2em}
    \caption{The WiReSens Toolkit web-based programming interface wirelessly records and displays pressure data from sensing devices (D) in real time, with panels to replay recordings (A), configure wireless protocols (B), and configure the devices themselves (C). The embroidered matt is configured to read only from the top-left quadrant and send readings via WiFi.}
  \label{fig:implementation}
\end{figure*}

\section{The WiReSens Toolkit}

The WiReSens Toolkit consists of a zero-potential readout circuit for interfacing with resistive sensing arrays and a Python-backed UI for configuring sensing devices, managing data recording and playback, and visualizing pressure data. We begin with a walkthrough to illustrate the core functionalities offered by our platform followed by the implementation details behind its development. 

\subsection{Walkthrough}

This walkthrough illustrates an example of setting up a tactile sensing system using a digitally embroidered 32×32 node resistive sensor and the WiReSens Toolkit.  While this example uses a specific fabrication method, we reemphasize that the toolkit is agnostic to sensor construction and supports resistive sensors produced through a wide range of manufacturing techniques.

\paragraph{\textbf{Hardware Setup}} 
To begin, the user connects the horizontal and vertical electrodes of their sensor to one of the zero-potential scanning readout circuits available in large (32×32) and small (16×16) configurations (Fig. \ref{fig:implementation}D). The large circuit interfaces with the sensor via ribbon cables and the small circuit does so via flex cables. In this case, the user selects the large circuit and connects the ESP32 microcontroller which controls it to their laptop via USB. The user then uploads the Arduino-based firmware, which handles sensor readout and communication. This upload is a one-time step—subsequent configuration is performed entirely through the web GUI.

\paragraph{\textbf{Configuring the Readout Area}}  
With the hardware connected, the user starts the Python backend and opens the web GUI in their browser. Using the Device Manager, they add a new device and access the configuration panel (Fig.~\ref{fig:implementation}C). The user decides they only need the top-left portion of the 32×32 array for their application, so they configure a smaller readout area of 16x16 nodes to improve the effective frame rate. A graphical interface allows intuitive selection of rows and columns, dynamically highlighting the chosen electrodes in green and providing a clear preview of the active sensing region. 

\paragraph{\textbf{Setting up Wireless Communication}}  
Next, the user configures wireless communication. The toolkit supports multiple protocols, including USB-serial, Wi-Fi, Bluetooth Low Energy (BLE), and ESP-NOW (Espressif’s peer-to-peer communication protocol~\cite{espressif_esp32_espnow}). The user selects Wi-Fi for its low latency and uses the wireless configuration panel (Fig.~\ref{fig:implementation}B) to enter parameters such as the laptop’s IP address and the network credentials.

\paragraph{\textbf{Programming the Device}}  
Once satisfied with the configuration, the user clicks ``Program Device'' to send the updated settings to the MCU via USB. If adjustments are needed later, the configuration can be updated directly through the web GUI, allowing for rapid iteration. The user then unplugs the device, connects a battery, and can begin using the device for their application.

\paragraph{\textbf{Recording, Visualization, and Playback}} 
By clicking ``Record,'' the user begins data acquisition across all configured devices, using their selected communication protocols and readout areas. They've also written a custom method for tracking touch points in Python, which hooks into the toolkit backend and runs concurrently with recording and visualization. Real-time pressure data is displayed as a heatmap of square ``sensing pixels'' representing the active readout regions. Users can reposition sensing pixels via drag-and-drop during recording to support flexible sensor layouts, in addition to saving these layouts and uploading background images for contextual visualization. Data is stored locally in HDF5 format, and a built-in replay feature enables playback with adjustable timestamps and speed controls (Fig. \ref{fig:implementation}A).

\begin{figure*}
  \includegraphics[width=\linewidth]{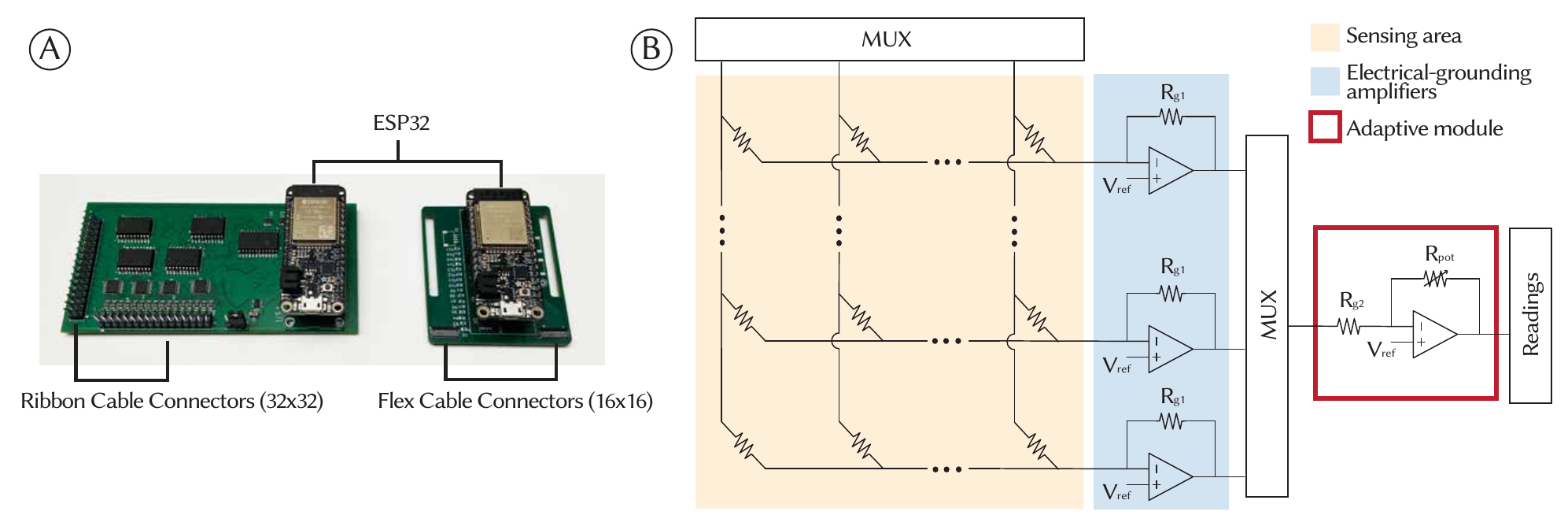}
  \vspace{-2em}
    \caption{(A) Adaptive zero-potential readout circuit open-sourced by WiReSens Toolkit in large (left) and small (right) sizes (B) Schematic of general zero potential readout circuit (left) with additional opamp and digital potentiometer for adaptivity (red).}
  \label{fig:circuit}
\end{figure*}

\paragraph{\textbf{Turning on low-power mode}} Aware of Wi-Fi’s energy demands, the user enables power-saving mode through the configuration panel (Fig. \ref{fig:implementation}C) to extend device lifetime during battery powered use. Using a Python utility which takes their pressure recording file as input, they obtain optimized intermittent transmission parameters that balance power efficiency with sensing quality for their specific application, and input these values into the configuration panel. They reconnect the MCU to their laptop using a USB cable and click the "Program Device" button, reconfiguring the device to operate in the low-power mode.

\paragraph{\textbf{Sensitivity Calibration}}  
During playback, the user notices that the sensor is less responsive than desired. By clicking ``Calibrate,'' they initiate an automatic gain calibration routine that optimizes voltage output range for the application at hand. During the 10-second calibration period, the user engages the tactile sensor in a way that reflects the range of pressures it would experience under normal operation. At the end of this interval, the firmware sets the amplification gain based on the recorded range of voltages, improving the sensor’s responsiveness for future interactions.

% \subsection{}
% Users configure their systems primarily through a JSON file which contains a variety of options for wireless communication, interactive visualization, and sensor readout. An overview of these options is available in our GitHub repository \footnote{\href{url-placeholder}{GitHub placeholder, code in supplemental materials for anonymity}}. Our incorporation of JSON promotes the development of tactile sensing systems which can change with minimal effort --- after a tactile sensing device has been flashed with initial base firmware (Fig.\ref{fig:implementation}B), users can simply update the file (Fig.\ref{fig:implementation}A) and pass it to the device (Fig.\ref{fig:implementation}C) to switch communication protocols, change the area of readout, increase sensor sensitivity, and much more

\subsection{Implementation} 

Here we provide further implementation details behind the toolkit's functionalities. 

\paragraph{\textbf{Web GUI and Python Library}}

The WiReSens Toolkit web GUI provides a visual programming interface for control of the toolkit's functionalities. It is implemented via NextJS and communicates with a locally running Python Flask backend through WebSockets. The Python backend handles data serialization for each device in a separate thread. It also provides a library with utility functions for analyzing sensor recordings, optimizing parameters for low power consumption, and executing custom methods on sensor data in real time. 

\paragraph{\textbf{Adaptive Readout Circuit}} The WiReSens Toolkit provides two open source zero-potential readout circuit PCBs which can be manufactured for \$90.26 and \$140.92 USD, respectively (Fig. \ref{fig:circuit}A). In this architecture, each electrode layer is connected to a digital multiplexer (MUX). To measure resistance at a specific node, the ESP32 controls one MUX to ground an electrode in one layer while the other MUX selects an electrode in the opposing layer for measurement. The voltage at this electrode is passed through an operational amplifier (LMV324, Texas Instruments) and digitized by the ESP32’s Analog-to-Digital Converter (ADC) at a continuous sampling rate of 75 kHz using direct memory access. To allow tuning of tactile sensor sensitivity and avoid the problems of ADC underutilization and op-amp output saturation,  we introduce an additional operational amplifier (TLV9152, Texas Instruments) in inverting configuration with variable gain (Fig. \ref{fig:circuit}B). The gain is controlled by a 50 k\(\Omega\) digital potentiometer (MCP4018, Microchip), denoted as \(R_\text{pot}\).

\paragraph{\textbf{Low-Power Mode}} To enable low-power operation modes, we employ intermittent sending, conserving power by minimizing data transmission for signals that are easy to predict or unchanging. The MCU predicts the voltage at a sensor node before reading it using the window-based time series forecasting equation:

\begin{equation}
v[n+1] = v[n] + \frac{1}{p} \cdot \left( v[n] - v[n-1] \right)
\label{eq:predictAlgo}
\end{equation}

 where \(v[n]\) represents the ADC voltage reading at time step \(n\) and \(p\) is a parameter controlling how responsive the prediction is to recent shifts in voltage. The ESP32 will only send the data packet if the average error between all readings and predictions is below some user defined threshold \textit{d}. If the data packet is not sent, the Python library will predict its value using the same forecasting method as the ESP32 (\ref{eq:predictAlgo}), ensuring that the average error across each data packet remains below the specified threshold \textit{d}. This is modeled after the method proposed by \citet{Suryavansh2021}, which was chosen due to its low computational complexity and memory requirements, making it ideal for implementation on a MCU.

 To help select the parameter \textit{p} and error threshold  \textit{d}, we provide a utility function which takes any tactile recording as input and performs a grid-based optimization. The optimization minimizes an objective function based on the normalized root mean square error (NRMSE) of the predicted tactile frames \(E\) and the percentage of transmitted data \(r\):
\begin{equation}
\alpha * E + (1- \alpha) * r,
\end{equation}

where \(\alpha\) is an adjustable parameter managing the trade-off between accuracy and transmission percentage. The script generates a visualization of the objective function, allowing users to refine and adjust the search space for further optimization.

\paragraph{\textbf{Automatic Sensitivity Calibration}} The automatic calibration sequence sets the resistance \(R_\text{pot}\) that best optimizes the tactile sensor's range of voltage output for a particular application. Upon start, the MCU scans the sensor array using the minimum possible value of \(R_{\text{pot}}\) for a configurable duration (defaulting to 10 seconds) and keeps track of the minimum sensor output voltages during this time. At the end of the calibration duration, the method then calculates the average of these minimum sensor output voltages \(V_{\text{min}}\) and determines the value of \(R_{\text{pot}}\) that will make the output of the opamp in the adaptive module \(V_{\text{out}}\) equal to 0 volts according to equation \ref{eq:vout}, obtained from analysis of the readout circuit topology in Fig. \ref{fig:circuit}, where the denominator \(264 \Omega\) represents the smallest detectable resistance by the readout circuit (as set by \(V_{\text{ref}} = 1.8V\) and \(R_{g1} = 220 \Omega\) in our implementation).

\begin{equation}
V_{\text{out}} = V_{\text{ref}} - \frac{R_{\text{pot}}}{264 \Omega} \times (V_{\text{ref}} - V_{\text{min}}).
\label{eq:vout}
\end{equation}

This effectively maps the minimum pressure applied to the maximum sensor output (\(V_{\text{ref}}\) Volts) and the maximum
pressure applied to the minimum sensor readout (0 Volts), ensuring that the full range of the ADC is used and pressure resolution is maximized.

\begin{figure}
  \includegraphics[width=\linewidth]{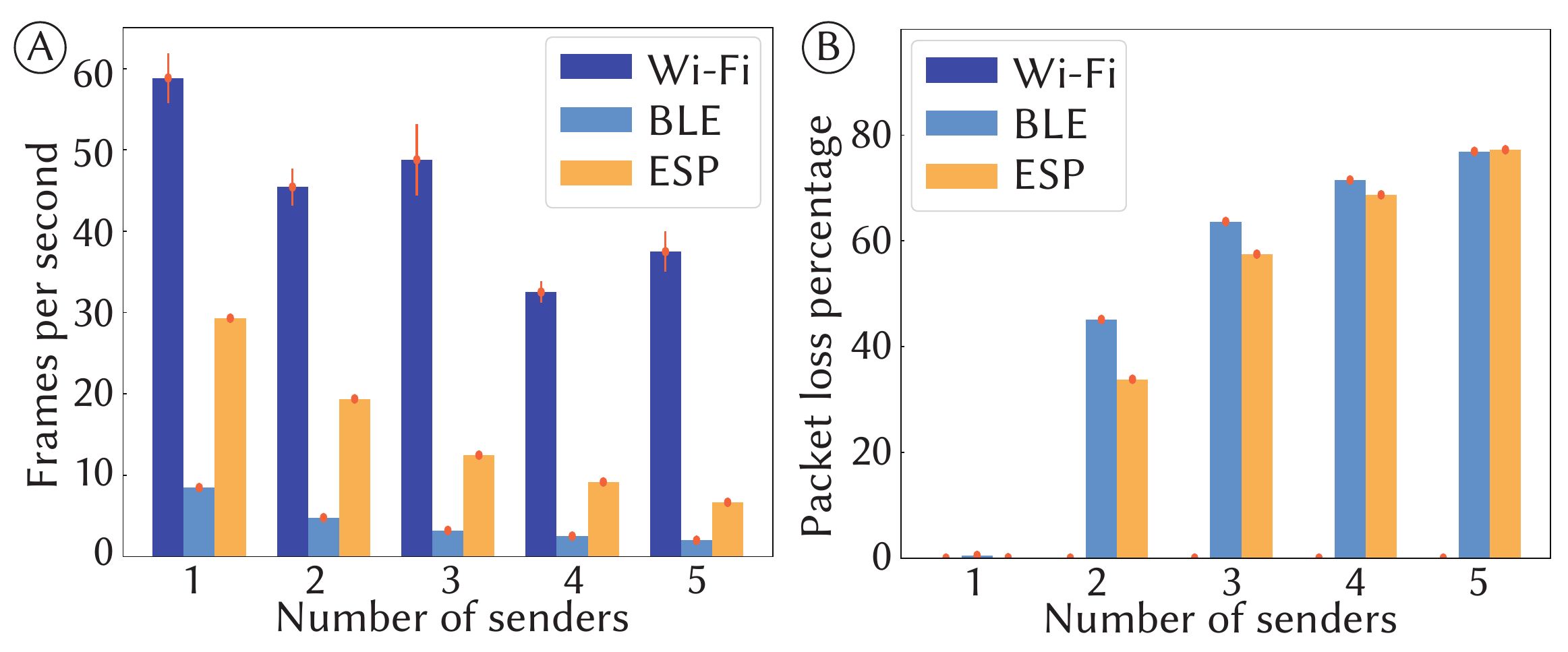}
  \vspace{-2em}
    \caption{Average throughput (A) and average percentage of packets lost (B) per sender during multi-sender wireless communication using Wi-Fi, BLE, and ESP-NOW.}
  % \caption{Multi-sender wireless communication performance for Wi-Fi, BLE, and ESP-NOW. (A) Average throughput per sender in frames per second, where one frame is equivalent to 1024 sensor readings or 2099 bytes of data and (B) Average percentage of packets lost per sender. All data is collected using our readout circuit as the sending device and a standard laptop computer as the receiving device.}
  \label{fig:communication}
\end{figure}

\begin{figure*}
  \includegraphics[width=\linewidth]{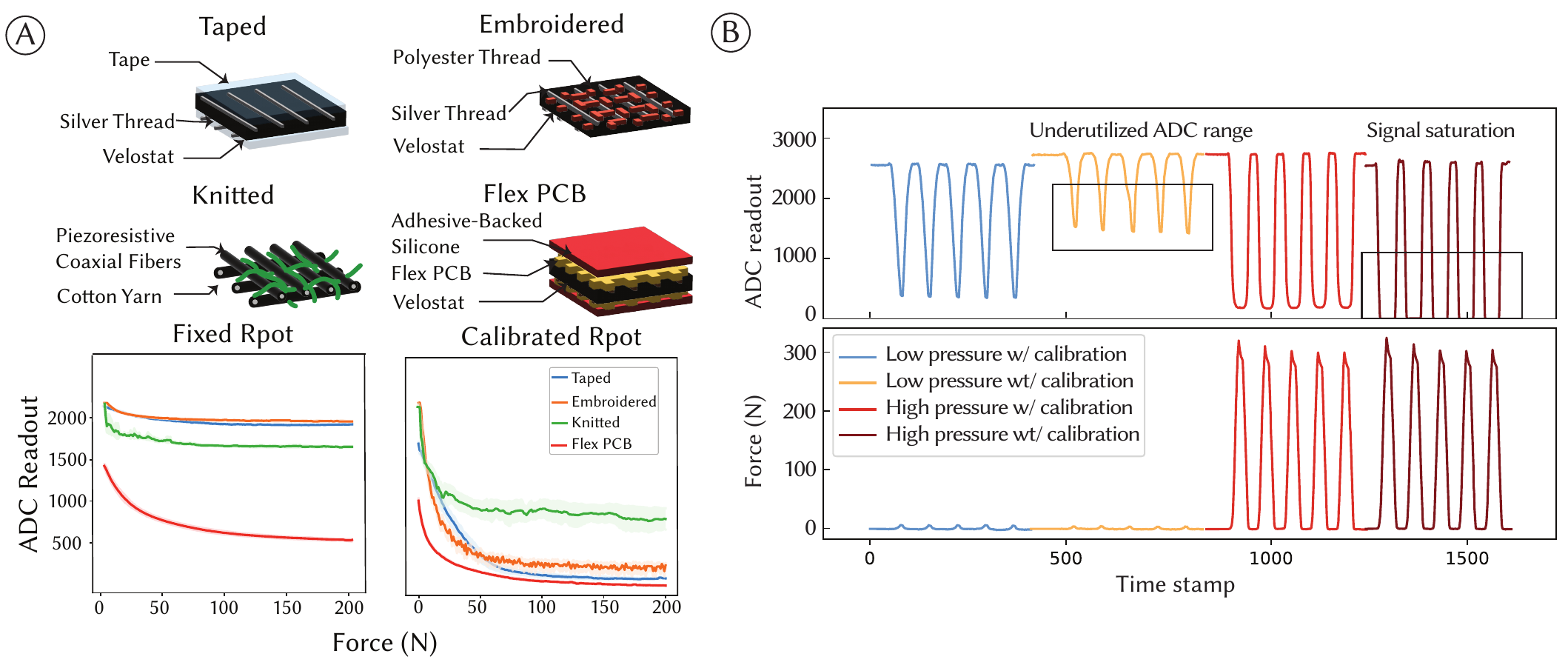}
  \vspace{-2em}
  \caption{WiReSens Toolkit calibrates readout for different sensors and application scenarios (A) Average and standard deviation of ADC readout for four different resistive pressure sensors during cyclical force tests, before and after calibration. (B) Average ADC readout during low and high-pressure application cycles, before and after calibration. Low-pressure calibration maximizes pressure resolution (blue and yellow curves) and High-pressure calibration avoids saturation (red and brown curves).}
  \label{fig:pot_data}
\end{figure*}

\section{Toolkit Evaluation}

In this section we offer a technical and user evaluation of the WiReSens Toolkit, focusing on speed and reliability of the multi-sender  wireless communication, automatic calibration performance,  power-reduction capabilities, and usability of the web-GUI interface. 
% during wireless operation. 

\subsection{Wireless Communication}

To assess the scalability of the wireless communication abstractions in WiReSens Toolkit (including Wi-Fi, BLE, and ESP-NOW), we programmed 1 to 5 devices to wirelessly send tactile sensing data from a 32x32 sensing matrix to a laptop running the Python backend. Each protocol and sender combination was tested for three minutes, repeated three times.  Fig. \ref{fig:communication} reports the average and standard deviation of throughput (in frames per second) and packet loss percentage per sender across all three tests. Here, one frame corresponds to approximately 2099 bytes.

% \begin{figure*}
%   \includegraphics[width=\linewidth]{figure/fabexperiment.pdf}
%   \vspace{-2em}
%   \caption{}
%   \label{fig:pot_fab}
% \end{figure*}
% \begin{equation}
%     \left(\frac{246 \, \text{bytes}}{\text{packet}}\right) \times \left(\frac{1024 \, \text{readings}}{\text{frame}} \cdot \frac{1 \, \text{packet}}{120 \, \text{readings}}\right) \ = 2099 \ \text{bytes}.
% \end{equation}

Throughput generally decreases and packet loss increases as the number of senders grows across all protocols due to limited bandwidth, though Wi-Fi showed exceptions to this trend that we attribute to network variability. Wi-Fi consistently achieved the highest performance, with over 60 fps and 0\% packet loss for a single sender, and over 30 fps with 0\% loss even with five senders. The narrower bandwidths of BLE \cite{bluetooth42} and ESP-NOW \cite{espressif_esp32_espnow} lead to more packet collisions during multi-sender, high-throughput scenarios, resulting in increased packet loss. However, the WiReSens Toolkit allows users to manage the trade-off between throughput and reliability by adjusting transmission frequency. For instance, with three senders, BLE achieved under 1\% packet loss with a 157 ms delay (0.75 fps per sender), and ESP-NOW achieved similar reliability with only a 10 ms delay (11.76 fps per sender). These findings suggest that BLE and ESP-NOW can serve as viable alternatives to Wi-Fi for users prioritizing lower power consumption or extended range, without significantly compromising data integrity.

\subsection{Sensor Adaptivity}

Here we evaluate the effectiveness of our calibration algorithm in two scenarios: adapting to different sensor fabrication techniques and different application contexts. Both experiments used a mechanical tester (Shimadzu AGX-V2) to apply an adjustable normal force over an area of 9 \(cm^{2}\) to a resistive sensor.

\paragraph{\textbf{Calibrating for Different Sensors}}  
We fabricated four resistive matrix-based pressure sensors using established methods from prior work: taping copper threads \cite{sundaram2019learning}, digital embroidery \cite{AignerEmbroidery2020}, machine knitting \cite{luo2021learning}, and flexible printed circuit board (FPCB) printing \cite{murphy2025fits}. For each sensor, we applied a cyclic force to a 3×4 node region at its center for five cycles, with the digital potentiometer's resistance (\(R_{\text{pot}}\)) set to \(787 \Omega\). Fig. \ref{fig:pot_data}A presents the mean and standard deviation of the ADC readout within the actuated region as a function of applied force for each fabrication method. Due to the inverting topology of the op-amp, the ADC readout is inversely correlated with force, exhibiting two approximately linear regions. Notably, the taped, embroidered, and knitted sensors display a significantly smaller overall change in ADC readout across the force range when using the default resistance, thereby reducing their effective pressure resolution.  

Next, we applied our auto-calibration procedure while maintaining a 200N applied force, yielding calibrated \(R_{\text{pot}}\) values of \(8661 \Omega\), \(4724 \Omega\), \(9055 \Omega\), and \(1181 \Omega\) for the taped, knitted, embroidered, and FPCB sensors, respectively. Repeating the cyclic force test, we observed enhanced pressure resolution in all four sensors post-calibration, as evidenced by the increased ADC output range over the same applied force range (Fig. \ref{fig:pot_data}A).

\paragraph{\textbf{Calibrating for Different Applications}} We then applied low and high-pressure sequences to the same 3x4 node region of the digitally embroidered sensor during a calibration period of 5 minutes and recorded the calibrated resistance values: \(R_{\text{pot}} = 14062.5 \Omega\) for the low pressure and \(R_{\text{pot}} = 8984.37 \Omega\) for the high pressure. We then repeated each pressure sequence twice:  once with  \(R_{\text{pot}} = 14062.5 \Omega\) and once with \(R_{\text{pot}} = 8984.37 \Omega\). Fig. \ref{fig:pot_data}B presents the applied force and average ADC readout over time for each sequence. When calibrated, the ADC readout range expanded for low-pressure scenarios, enhancing resolution, while high-pressure calibration prevented 0V saturation, allowing sensitivity up to 250 N.

\begin{figure*}
  \includegraphics[width=\linewidth]{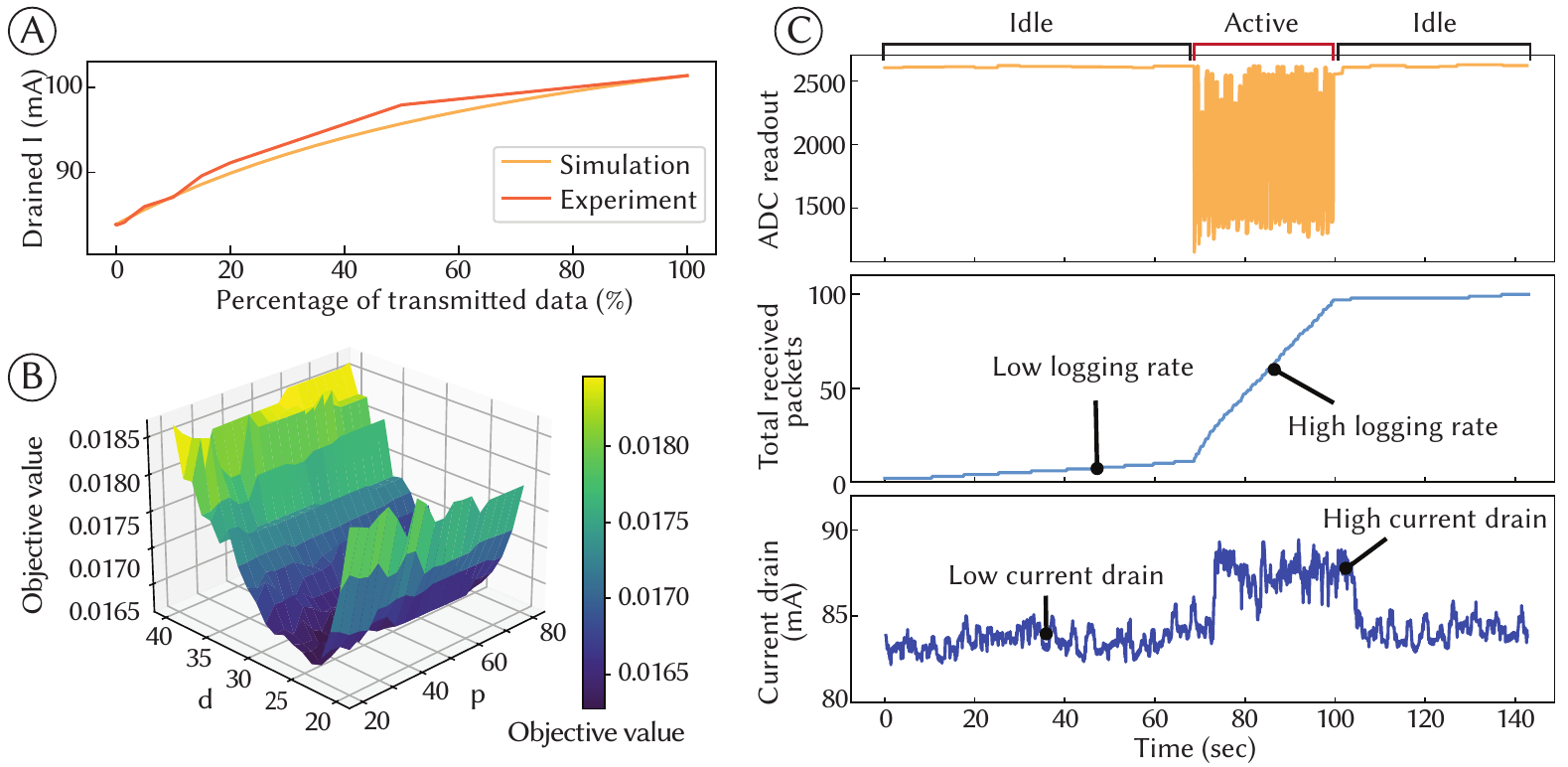}
  \vspace{-1.5em}
  \caption{Characterization of intermittent sending performance. (A) Simulated and observed average current draw (mA) from a tactile sensing device as a function of the percentage of data transmitted during BLE operations. (B) Visualization of objective function used to select intermittent sending parameters. (C) Average ADC readout, received packet count, and current draw over time during periods of no applied pressure (Idle) and repeated pressure (Active). Low current drain indicates power is saved during Idle periods. }
  \label{fig:intermittent}
\end{figure*}

\subsection{Wireless Power Reduction}

We first investigate how packet transmission rate affects power consumption in a tactile sensing device using the BLE protocol. We use an abstraction where the device alternates between two states: sending data (state \(A\)) or idle (state \(B\)). Then, the average current drained during device operation is calculated as
\begin{equation}
I_{A}*t_{A} + I_{B}*(1-t_{A}),
\end{equation}

where \(I_{A}\) and \(I_{B}\) is the average current draw during states A and B, respectively, and \(t_{A}\) is the time percentage state in state A. \(I_{B}\) is measured while the MCU samples but does not send pressure data, and \(I_{A}\) is estimated by subtracting \(I_{B}\) from the measured current while data is sent continuously. Using a Nordic Semiconductor Power Profiler, we log the current (mA) from a 3.3V 1200mAH battery over three minutes to obtain these averages. 

We then simulate how the percentage of transmitted data changes \(t_{A}\), and thus the average current draw of the device. We validate this simulation by fixing packet transmission frequency at 10 different values between 0 and 100\% and recording the average current draw. Fig. \ref{fig:intermittent}A shows the simulated and observed trends, with results indicating that sending ~1\% of packets could extend the device lifetime by over 20\% for BLE. Using similar calculations with data collected under Wi-Fi operation, we estimate this could increase to 42 \% longer battery life, as Wi-Fi operation is generally more power-intensive than BLE (we measured 152.47 mA vs 101.31 mA for BLE).

To further validate the intermittent sending performance, we use a mechanical tester to apply a series of repeated presses to a 32x32 tactile sensor during intermittent operation. Using the Python utility, we optimize the transmission parameters \textit{p} and \textit{d} (\(p=29\) and \(d=26\), Fig. \ref{fig:intermittent}B) and use them to program the device for the live pressure test. The results (Fig. \ref{fig:intermittent}C), indicate that when the sensor is inactive, packet transmission is minimal with low average current drain compared to when the sensor is active. The optimized values achieved packet transmission just under 5\% with a NRMSE of around 0.012. With a 1200mAH battery, we estimate that this would extend the device's lifetime by over 2 hours, while ensuring the predicted tactile frames differed from the ground truth frames by approximately only 1.2\% of the full-scale range.

% These results demonstrate that our intermittent sending method can prolong the operational life of resistive tactile sensing devices while maintaining their functionality. 

\subsection{User Study}

In our user study, we assessed whether first-time users found the WiReSens Toolkit easy to use and to what extent it simplifies prototyping resistive tactile sensing systems. Participants completed three tasks: (1) programming a new sensor, (2) calibrating a sensor, and (3) reproducing recorded pressure data. This study was approved by our institutional review board.  

\begin{figure}
    \centering
    \includegraphics[width=1\linewidth]{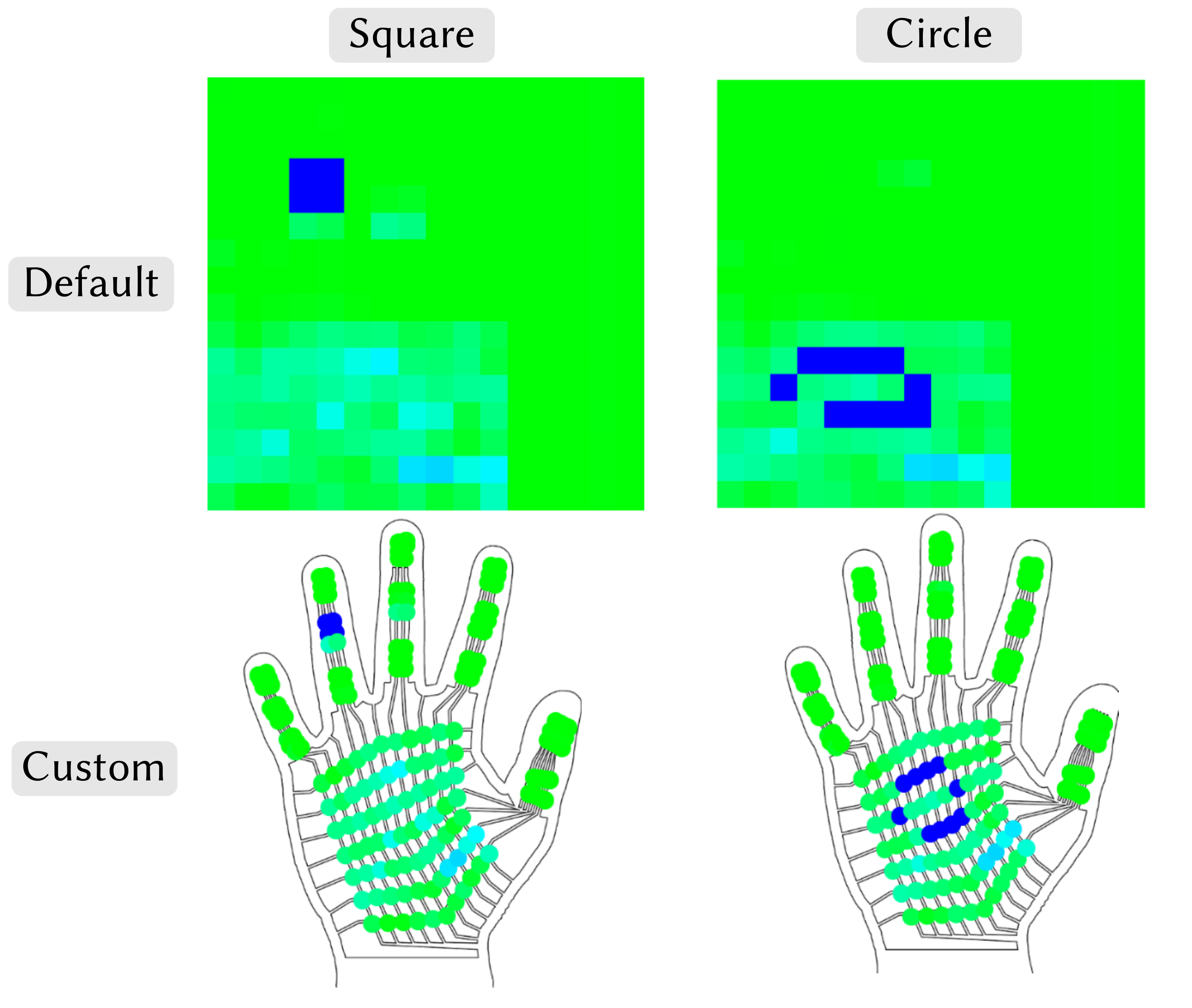}
    \caption{Study participants are asked to reproduce one square press and one circle press from either a default or custom visualization.}
    \label{fig:presses}
\end{figure}

\paragraph{\textbf{Participants}}
We recruited 11 participants (4 female, 7 male) aged 19–32 (M = 22.64, SD = 3.87) through university mailing lists and word of mouth. Only two had prior experience with resistive pressure sensors, and none had used any part of the WiReSens Toolkit before.  

\paragraph{\textbf{Task 1: Programming a Device}}  
To evaluate the toolkit’s learnability and usability, participants were tasked with using it to program a new device. They first received an overview of the resistive pressure sensing principle and were introduced to a 32×32 digitally embroidered sensor. The study moderator then demonstrated how to configure a device to read and record pressure data over Bluetooth using the web GUI. After the demonstration, participants could ask questions before attempting the task independently. They were asked to configure the device to read and record data from a new sensor—an 8×24 shoe sole—without guidance. They could reconfigure the device until they were confident in their setup. We recorded the time taken and the configured readout area. At the end of the study, participants completed a System Usability Scale (SUS) survey \cite{SUS2009} to evaluate their experience.  

\paragraph{\textbf{Task 2: Calibrating a Device}}  
Next, we assessed the extent to which our auto-calibration procedure simplifies the calibration process. The study moderator preconfigured a new device to read from a 3×2 node area of a flex PCB sensor while a constant force of 3.0 kg was applied. Participants were given background information on the sensitivity calibration challenges of resistive sensors and were then instructed to manually adjust the digital potentiometer value through the web GUI to find the minimum amplification such that all six nodes displayed an ADC reading of zero. We recorded the time taken for this task and the user-selected value. The study moderator then asked the participant to perform the same calibration using the "Calibrate" button, which ran for a fixed duration of 10 seconds, and recorded the auto-calibrated potentiometer value.

\paragraph{\textbf{Task 3: Interpreting Visualized Pressure Data}}  
Finally, we examined whether the toolkit's customized visualizations improve spatial reasoning about tactile data. Participants were shown two WiReSens Toolkit visualizations of presses on a sensing glove—one created with a circular object and one with a square object. Each visualization was presented in one of two formats: either a custom layout configured to match the glove’s shape or a default square layout (Fig. \ref{fig:presses}). The order of presentation was counterbalanced across participants. Participants were asked to reproduce the two presses using the objects and could adjust their attempt until they felt confident. We recorded the time taken and asked them to complete a survey assessing their confidence and perceived difficulty during each task on a Likert scale rating from 1 (Not Difficult/Not Confident) to 7 (Very Difficult/Very Confident).

\paragraph{\textbf{Results}}

\begin{figure}
    \centering
    \includegraphics[width=1\linewidth]{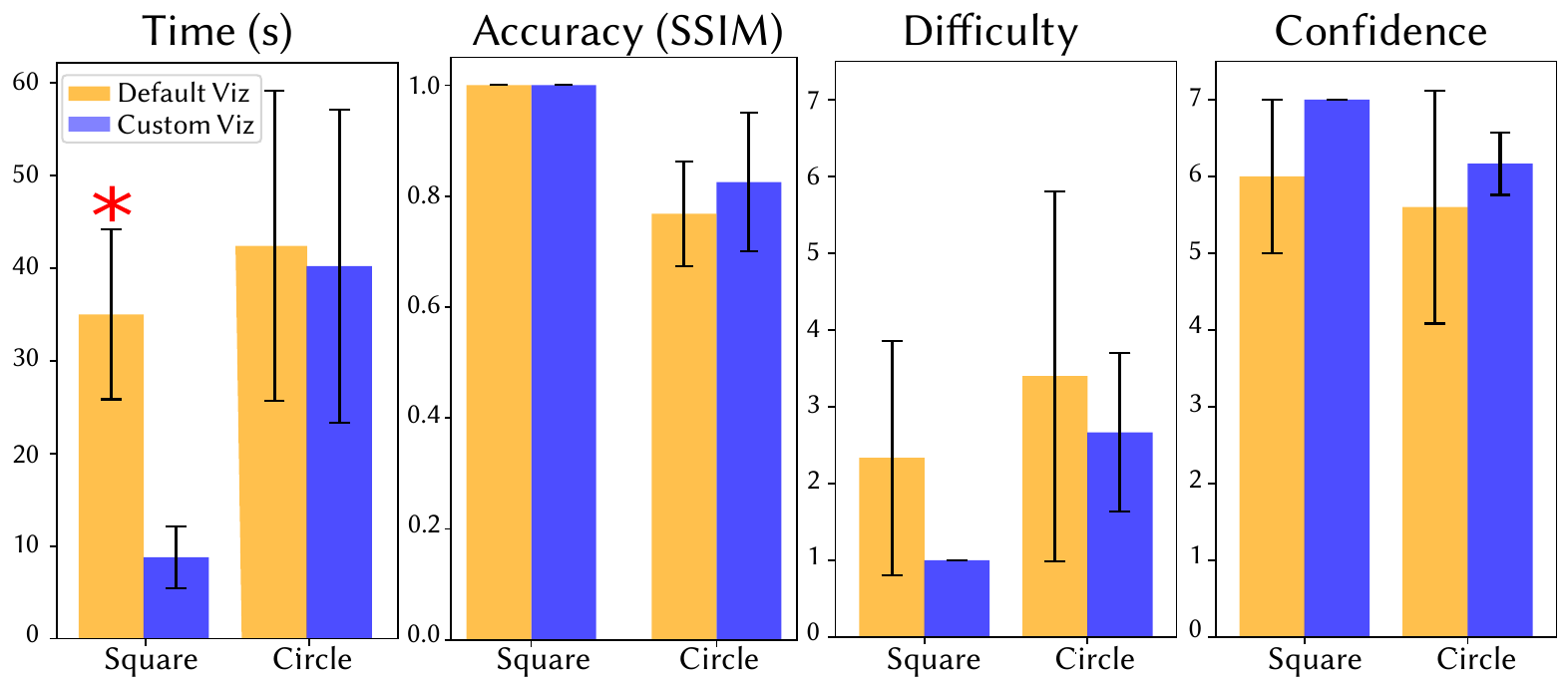}
    \caption{Mean and standard deviation of time (lower is better), accuracy (higher is better), difficulty (lower is better), and confidence (higher is better) for square and circle press tasks. Results are shown for default (orange) and custom (blue) visualizations.}
    \label{fig:studyresults}
\end{figure}

\begin{figure*}
  \includegraphics[width=\linewidth]{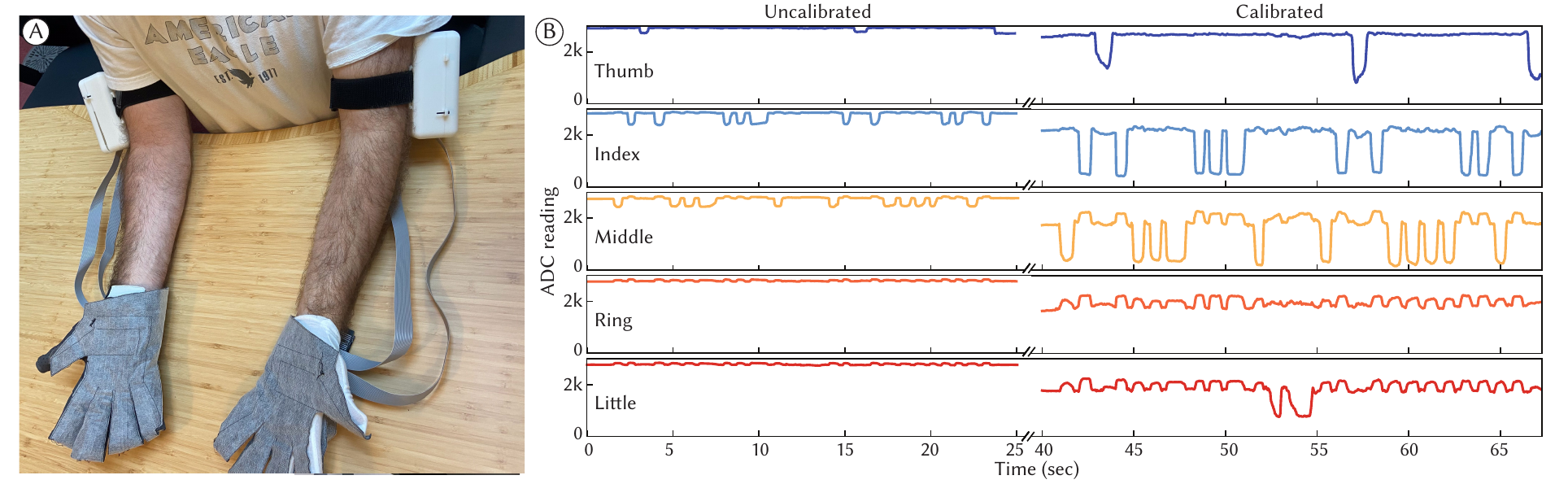}
  \vspace{-1em}
  \caption{Wireless Musical Gloves: (A) Depiction of tactile sensing array, with readout circuit affixed to the arm via velcro straps (B) Average ADC readout in the Thumb, Index, Middle, Ring, and Little finger regions of one glove during "Mary Had a Little Lamb", before and after calibration, showing higher sensitivity after calibration.}
  \label{fig:glove}
\end{figure*}

On average, participants took less than 5 minutes (M = 4.15, S.D = 1.71) to program a new sensor, achieving an average readout area accuracy of 96\% (S.D = 4.3\%) on their first attempt. The average SUS rating was 82.5 (S.D = 9.68), considered excellent and falling in the highest quartile of usability scores \cite{Bangor29072008}. Open-ended feedback further highlights the toolkit’s novice-friendly nature, as reflected in P11’s comment: \textit{“It was very cool to work with hardware integrated with the toolkit software because it made the learning process more smooth and approachable. It was also beginner-friendly, and I could see myself using this for a project.”}

% They primarily used one of two methods to configure the readout area: either directly interacting with the sensor to observe which areas of the visualization responded to pressure or analyzing how the electrodes connected from the sensor to the readout circuit. Those who relied on direct interaction often configured the readout area with one fewer electrode on either the row or column side, likely due to weaker signals near the sensor edges, indicating that participants were already using the interface to filter out unnecessary information. 

The toolkit expedites the tedious task of calibration. While all participants found the optimal \(R_{\text{pot}}\) value through manual calibration, this took 2.1 minutes on average (S.D = 1.15). The auto-calibration always found the same user-identified optimal gain, but with a fixed calibration duration of 10 seconds, was over 10x faster. Three participants mention the auto-calibration as their favorite feature in open-ended feedback - P2 notes: \textit{"I liked the auto-calibration tool because it seemed quite tedious to find the resistance yourself."}

Finally, the WiReSens Toolkit's custom visualization improves spatial reasoning of tactile data. We evaluated its effectiveness by analyzing participants' task completion time, Likert scale ratings for difficulty and confidence, and accuracy of reproduced presses using the Structural Similarity Index Measure (SSIM) between binary thresholded participant and ground truth presses (Fig. \ref{fig:studyresults}). Participants using the custom visualization in the square press task took significantly less time (M = 8.79, S.D = 3.32, Welch’s t-test p < 0.05) and reported less difficulty and greater confidence on average. In the more challenging circle press task, participants using the custom visualization also performed better, taking less time, reproducing presses more accurately, and expressing higher confidence than those using the default visualization, on average. Six participants highlighted the visualization tools as their favorite feature, with P8 stating: \textit{“The layout is my favorite feature, the ability to hide and reconfigure layouts to match the shape of the object.”
}

\section{Example Applications}

\label{section:application}

In this section, we demonstrate the utility of WiReSens Toolkit by using it to prototype a range of resistive tactile sensing systems. The tactile sensors were constructed with silver thread and Velostat, using digital embroidery \cite{AignerEmbroidery2020} and direct taping \cite{adafruit_firewalker}. Each application was prototyped in under 10 minutes.

\paragraph{\textbf{Musical Gloves}} Taking inspiration from gloves that offer new means of musical expression \cite{NIME20_62, orth1998musical}, we used WiReSens Toolkit to program two Wi-Fi-enabled tactile sensing gloves for playing a virtual piano, showcasing its potential for enhancing real-time, expressive musical performance. We recruited two users—one for the left-hand glove and one for the right-hand glove—to perform a piano duet both before and after calibration. The ADC readout in each finger region is shown in Fig. \ref{fig:glove} for one of the users. After calibration, each user reported feeling like their glove was more responsive to their applied pressure, which can be evidenced by the greater range of ADC readout shown in Fig. \ref{fig:glove}B. 

\begin{figure*}
  \includegraphics[width=\linewidth]{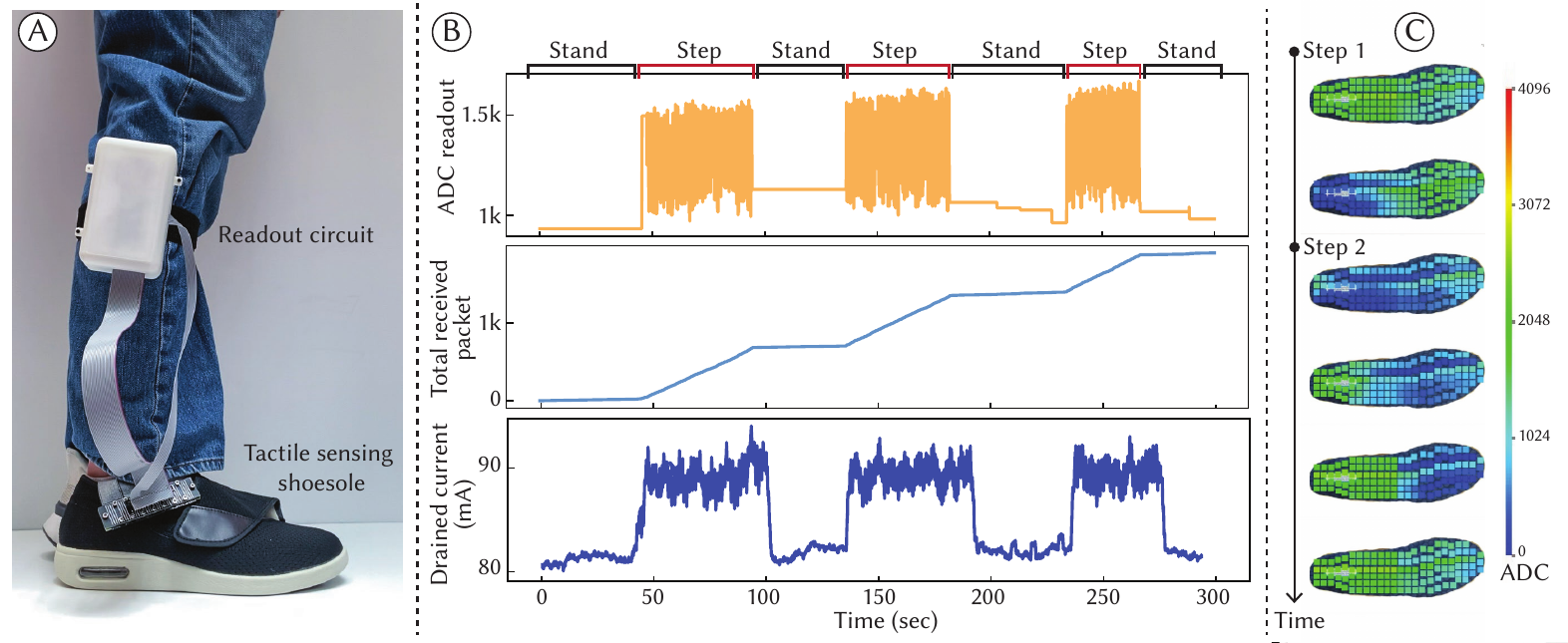}
  \vspace{-2em}
  \caption{Gait Monitoring Shoe Sole: (A) Tactile sensing shoe sole form factor and readout circuit. (B) Power efficient operation using intermittent sending, with reduced current drain during periods of standing still and (C) Customized visualization for gait monitoring and step tracking.}
  \label{fig:shoe sole}
\end{figure*}

\paragraph{\textbf{Gait Monitoring Shoe Soles}} Prior research has demonstrated interest in gait monitoring as a valuable tool for understanding overall health \citep{Liu2021Gait, MIRELMAN2019Parkinsons}. We programmed a BLE-enabled pressure monitoring shoe sole, recorded pressure data during 5 minutes of walking and standing still, and used the toolkit's interactive visualization to replay the wirelessly recorded gait. As shown in Fig. \ref{fig:shoe sole}, our device consumes considerably less power while the user is standing still than when they are stepping. The significant increase in packet transmission during the walking periods evidences that this is a direct result of our intermittent send algorithm ensuring efficient use of the onboard communication modules.

\begin{figure}
    \includegraphics[width=\linewidth]{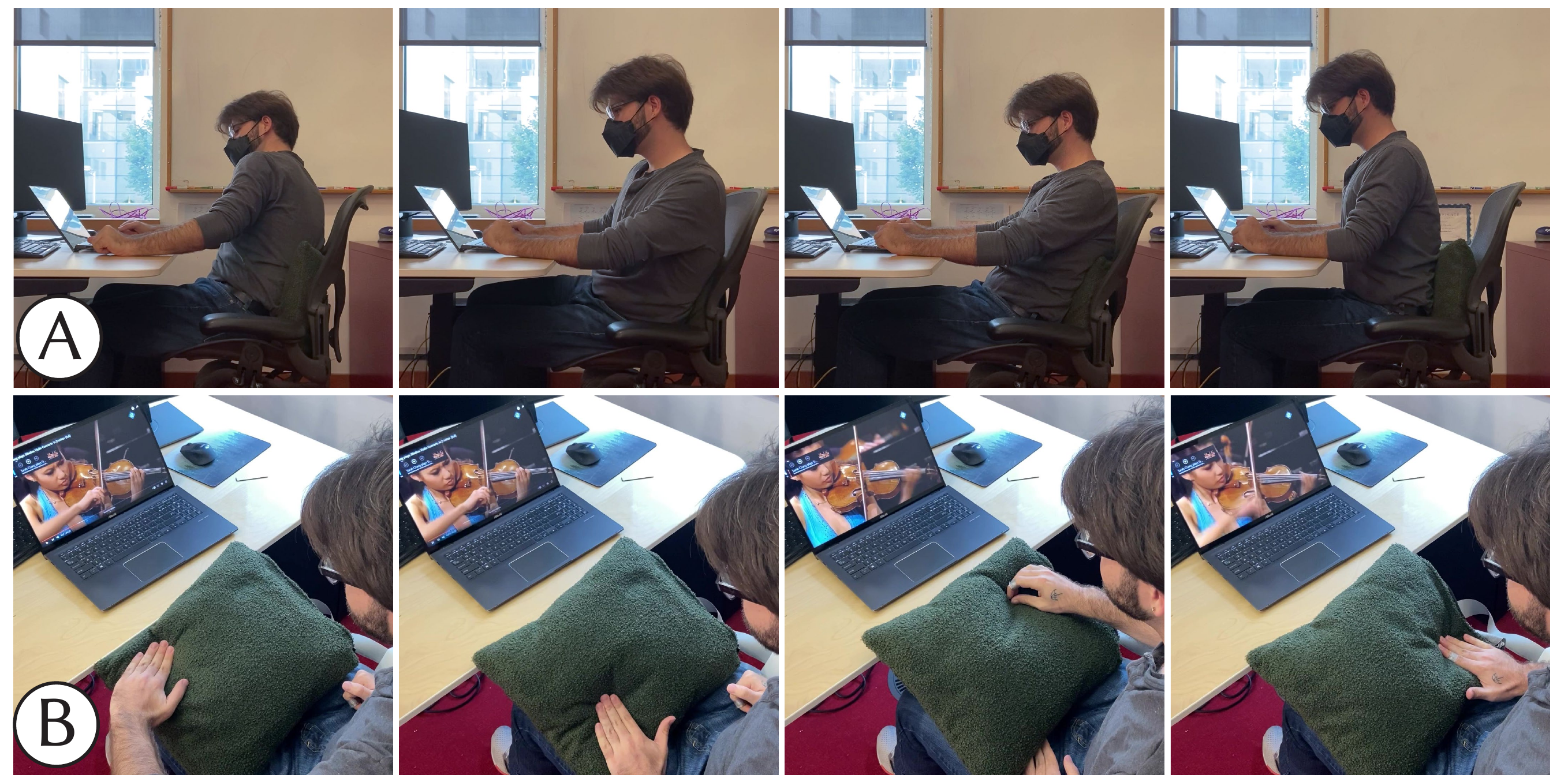}
  \vspace{-1em}
  \caption{Tactile Sensing Pillow: Used for (A) Posture monitoring, from left to right: leaning right, leaning left, leaning back, sitting up straight and (B) Remote Control, from left to right: Play, Pause, Volume Up, Volume Down.}
  \label{fig:pillow}
\end{figure}

\begin{figure}
  \includegraphics[width=\linewidth]{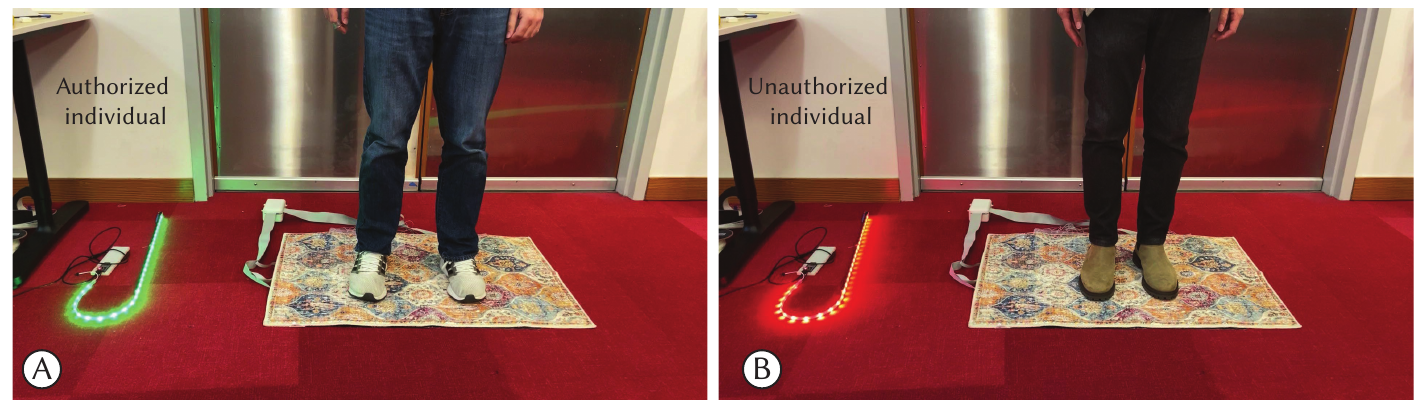}
  \vspace{-1em}
  \caption{Smart Home Welcome Mat: IoT lamp processes tactile sensing data from a welcome mat to turn green when (A) an authorized individual enters the home and red when (B) an unauthorized individual enters the home.}
  \label{fig:mat}
\end{figure}

\paragraph{\textbf{Other applications}} Because the sensitivity of readout can adaptively change during device operation, WiReSens Toolkit enables the development of tactile sensing devices that can be repurposed for different applications. We use this feature to develop a tactile sensing pillow that can serve both as a passive recording device (higher sensitivity) for posture monitoring and as an active input interface (lower sensitivity) for remote control of media playback (Fig. \ref{fig:pillow}). We also developed a smart home IoT welcome mat (Fig. \ref{fig:mat}). The system includes a tactile sensing mat and a separate ESP32-controlled lamp. The mat captures tactile data, which is wirelessly transmitted using ESP-NOW to another microcontroller responsible for turning the lamp on when an authorized user is standing on it. In future iterations, WiReSens Toolkit could be used to enable seamless integration with other smart home devices, such as doorbells or security cameras.

\section{Discussion}

In this section, we discuss the primary limitations of WiReSens Toolkit and provide insight into how they may be addressed. We further outline several avenues for future works that WiReSens Toolkit's contributions enable. 

\subsection{Limitations}  

While WiReSens Toolkit can be used in remote settings, portability may be limited by the fact that BLE and ESP-NOW communication are fundamentally lower-throughput, lower-range protocols \cite{commStudy} and backend support is restricted to desktop computers. This can be addressed by adding support for multiple receivers to balance the data load more effectively and implementing the backend within a mobile application. Additionally, we chose intermittent data transmission as a computationally efficient and intuitive means of power reduction that would not hinder sensing capability, but energy savings may be limited when compared to battery-free systems \citep{kortbeek2020bfree, kraemer2022battery}. Future work may explore ultra-low-power wake-on-motion sensing or energy harvesting techniques to extend operational lifetime without compromising sensing fidelity. Moreover, while the ESP32 microcontroller was chosen for accessibility, it imposes constraints on sample rate, power consumption, and form factor. Using ASIC or FPGA-based sensor readout could enhance the system’s ability to detect finer-grained or fast-changing pressures, such as vibrations, while reducing power consumption and size.

\subsection{Future Work}

While this work focuses on resistive sensors for tactile interactions, the WiReSens Toolkit can extend to other sensing modalities that rely on conductivity changes, such as thermistors for temperature, moisture sensors, chemiresistors for gas detection, and photoresistors for light intensity. We see this adaptability as enabling large-scale monitoring of multimodal physical phenomena beyond just touch. Additionally, future work could integrate on-device learning to dynamically optimize toolkit parameters for specific applications. For instance, in a wearable health monitoring application, the system could adjust sensor calibration in response to physiological changes like swelling or muscle tension, ensuring long-term accuracy. Finally, we envision this work advancing tactile sensing as a privacy-conscious and efficient alternative to vision-based systems, as resistive tactile sensing data is lower resolution and largely unidentifiable by humans. 

\section{Conclusion} 
We introduced the WiReSens Toolkit, designed to facilitate the creation of portable, adaptive, and long-lasting resistive-matrix-based tactile sensing devices. Our implementation supports variably-sized resistive sensor array readout, wireless communication via Wi-Fi, BLE, or ESP-NOW protocols, adaptive sensitivity through auto-calibration, and power efficiency through intermittent computing techniques. Additionally, it streamlines the logging and real-time custom visualization of tactile sensing data from multiple devices. We conducted a technical evaluation of WiReSens Toolkit's functionality, providing evidence of its ability to enable fast and robust multi-sender communication, maintain consistent sensor performance across pressure intensities, and ensure power-efficient operation. Application prototypes and a comprehensive user study further validate the interface as a usable and useful tool for developing interactive systems based on the resistive sensing principle. 

We believe WiReSens Toolkit will seamlessly integrate with ongoing advancements in digital manufacturing, wireless sensor networks, and artificial intelligence, empowering the next generation of tactile sensing technologies for deeper insights into human interactions with the physical world. 

\bibliographystyle{ACM-Reference-Format}
\bibliography{10_reference}

\end{document}